\documentclass[10pt]{book}
\usepackage{epsf,latexsym,sussp2004}

\newcommand{\half}{\mbox{\small $\frac{1}{2}$}}          
\newcommand{\msbar}{\mbox{\tiny $\overline{MS}$}}        

\def\lsim{\mathrel{\rlap{\lower4pt\hbox{\hskip1pt$\sim$}}
    \raise1pt\hbox{$<$}}}                
\def\gsim{\mathrel{\rlap{\lower4pt\hbox{\hskip1pt$\sim$}}
    \raise1pt\hbox{$>$}}}                

\def\3{\ss}

\newcommand{\Dd}[1]{\mbox{
  \parbox[b]{0cm}{$D$}\raisebox{1.7ex}{$\leftrightarrow$}$_{\!#1}$}}

\begin{document}

\setcounter{section}{1}
\setcounter{page}{1}

\headings{The Lattice Calculation of Moments of Structure Functions}
         {The Lattice Calculation of Moments of Structure Functions}
         {R. Horsley}{School of Physics\\The University of Edinburgh}


\section{Introduction}

\vspace{-4.50in}                                     
{\normalsize Edinburgh 2004/30}\\                    %
{\normalsize December 2004}    \\                    %
\vspace{3.80in}                                      %


Much of our knowledge about QCD and the structure of hadrons
(mainly nucleons) has been gained from Deep Inelastic Scattering (DIS)
experiments such as $eN \to eX$ or $\nu N \to \mu^- X$.
The (inclusive) cross sections are determined
by structure functions $F_1$ and $F_2$ when summing over beam and target
polarisations (and an additional $F_3$ when using neutrino beams), and
$g_1$, $g_2$ when both the beam and target are suitably polarised.
Structure functions are functions of the Bjorken variable $x$ and $Q^2$,
the large space-like momentum transfer from the lepton.
(Another class of structure functions -- the transversity $h_1$ --
can be measured in Drell-Yan processes or certain types of semi-inclusive
processes.)

As a direct theoretical computation of structure functions does not
seem to be possible, we must turn to the Wilson Operator Product Expansion
(OPE) which relates moments of structure functions to (nucleon) matrix
elements in a twist (ie operator [dimension - spin]) or Taylor expansion
in $1/Q^2$. So first defining bilinear quark operators
\begin{equation}
   {\cal O}_q^{\Gamma ; \mu_1\cdots\mu_n} =
       \overline{q} \Gamma^{\mu_1\cdots\mu_i}
                       i\Dd{}^{\mu_{i+1}} \cdots i\Dd{}^{\mu_n} q \,,
\end{equation}
where $q$ is taken to be either a $u$ or $d$ quark and $\Gamma$ is an
arbitrary Dirac gamma matrix
we have for the nucleon matrix elements, the Lorentz decompositions
($s^2 = - m_N^2$)
\begin{eqnarray}
   {1\over 2} \sum_s
   \langle \vec{p}, \vec{s}) |
         \widehat{\cal O}_q^{\gamma ; \{\mu_1 \cdots \mu_n\}}
                               | \vec{p}, \vec{s} \rangle
     \hspace*{-0.075in} &=& \hspace*{-0.075in} 2v^{(q)}_n
          \left[ p^{\mu_1} \cdots p^{\mu_n} - \mbox{tr} \right] \,,
                                         \nonumber  \\
   \langle \vec{p}, \vec{s} |
         \widehat{\cal O}_q^{\gamma\gamma_5 ; \{\sigma\mu_1 \cdots\mu_n\}}
                               | \vec{p}, \vec{s} \rangle
     \hspace*{-0.075in} &=& \hspace*{-0.075in} a^{(q)}_n
          \left[ s^{ \{ \sigma} p^{\mu_1} \cdots p^{\mu_n \}}
                                            - \mbox{tr} \right] \,,
                                         \nonumber  \\
   \langle \vec{p},\vec{s} |
         \widehat{\cal O}_q^{\gamma\gamma_5 ;
                             [  \sigma \{ \mu_1 ] \cdots \mu_n \} }
                             | \vec{p},\vec{s} \rangle
     \hspace*{-0.075in} &=& \hspace*{-0.075in} {n d^{(q)}_n \over{n+1}} 
          \left[ (s^{\sigma} p^{\{ \mu_1} - p^{\sigma} s^{\{ \mu_1} )
                            p^{\mu_2}\cdots p^{\mu_n \}} - \mbox{tr}
          \right] \,,
                                              \nonumber \\
   \langle \vec{p}, \vec{s} |
         \widehat{\cal O}_q^{\sigma\gamma_5;
             \sigma\{\mu_1 \cdots\mu_n\}}
                                       | \vec{p}, \vec{s} \rangle
     \hspace*{-0.075in} &=& \hspace*{-0.075in} {t^{(q)}_{n-1} \over m_N}
          \left[ (s^{\sigma} p^{\{ \mu_1} -  p^{\sigma} s^{\{ \mu_1} )
                            p^{\mu_2}\cdots p^{\mu_n \}} - \mbox{tr}
          \right] \,,
\end{eqnarray}
where the symmetrisation/anti-symmetrisation operations on the operator
indices also indicates that they are traceless (which gives them a
definite spin). $v_n$, $a_n$, $d_n$ and $t_n$ can be related to moments
of the structure functions. For example we have for $v_n$ and $F_2$
\begin{equation}
   \int_0^1 dx x^{n-2} F_2(x,Q^2) =
      {1\over 3} \sum_{f=u,d,g,\ldots}
          E^{(f)\msbar}_{F_2;n}( \mu^2/Q^2, g^{\msbar} )
            v_n^{(f)\msbar}(\mu) + O(1/Q^2) \,,
\label{f2sf}
\end{equation}
and similar relations hold between $g_1$ and $a_n$; $g_2$
and a linear combination of $a_n$ and $d_n$; $h_1$ and $t_n$.
Although the OPE gives $v_n$ from $F_1$ (or $F_2$) for $n = 2, 4, \ldots$;
$v_n$ from $F_3$ for $n = 3, 5 , \ldots$; $a_n$ from $g_1$ for
$n = 0, 2, \ldots$; $a_n$, $d_n$ from $g_2$ for $n = 2, 4, \ldots$, other
matrix elements can be determined form semi-exclusive experiments,
for example $a_1$ by measuring $\pi^{\pm}$ in the final state.

While the Wilson coefficients, $E^{\msbar}(1,g^{\msbar}(Q))$ are
known perturbatively (typically two to three loops) and determine
how the moments change with scale, the `inital condition' ie the
matrix element is non-perturbative in nature. The only known way
of determining them from QCD in a model independent way is via
Lattice Gauge Theory (LGT). In this talk we review our status 
(QCDSF and UKQCD Collaborations) of some aspects of these
determinations including some higher twist results.
A more general review may be found in (G\"ockeler \textit{et al} 2002a).
We shall also restrict ourselves here to forward matrix elements
(and so not consider the form factors and the more embracing
Generalised Parton Densities or GPDs). Determining moments
of structure functions is an active field of reseach at present,
see for example (Dolgov \textit{et al} 2002,
Guagnelli \textit{et al} 2004, Ohta \textit{et al} 2004).


\section{The Lattice Approach}


The lattice approach involves first Euclideanising the QCD action and
then discretising space-time with lattice spacing $a$. The path integral
then becomes a very high dimensional partition function, which is amenable
to Monte Carlo methods of statistical physics. This allows ratios of
three-point to two-point correlation functions to be defined,
\begin{equation}
   R_{\alpha\beta}(t,\tau; \vec{p})
      =       {\langle N_\alpha(t;\vec{p}) {\cal O}_q(\tau)
                   \overline{N}_\beta(0;\vec{p}) \rangle \over
                  \langle N(t;\vec{p}) \overline{N}(0;\vec{p}) \rangle}
      \propto \langle N_\alpha(\vec{p})|
                   \widehat{\cal O}_q |N_\beta(\vec{p}) \rangle \,,
\label{Rratio}
\end{equation}
where $N_\alpha$ is some suitable nucleon wavefunction (with Dirac 
index $\alpha$) such as
\begin{equation}
   N_{\alpha}(t;\vec{p}) = \sum_{\vec{x}} e^{-i\vec{p}\cdot\vec{x}}
                           \epsilon^{ijk} u^i_\alpha(\vec{x},t)
                           [ u^j_\beta(\vec{x},t) (C\gamma_5)_{\beta\gamma}
                             d^k_\gamma(\vec{x},t) ] \,.
\end{equation}
The proportionality holds for $0 \ll \tau \ll t \lsim \half N_T$
for a lattice of size $N_S^3\times N_T$.
There are two basic types of diagrams to compute in eq.~(\ref{Rratio}):
the first is a quark insertion in one of the nucleon quark lines
(`quark line connected'), while in the second type the operator
interacts only via gluon exchange with the nucleon (`quark line disconnected').
Due to gluon UV fluctations these latter diagrams are numerically difficult
to compute. However by considering the Non-Singlet, NS,
or ${\cal O}_u - {\cal O}_d$ operators, giving matrix elements
such as $v_{n;NS} = v_n^{(u)} - v_n^{(d)}$
in eq.~(\ref{f2sf}) then the $f = s$ and $g$ (gluon) terms cancel.
(For higher moments however, one might expect that sea effects
anyway are less significant as the integral is more weighted to $x \sim 1$.)
Although LQCD is in prinicple an `ab initio' calculation there
are, of course, several caveats. First our lattice `box' must be large 
enough to fit our correlation functions into. A continuum limit
$a \to 0$ must be taken. A chiral extrapolation must be made from
simulations often at the strange quark mass or larger down to the almost
massless $u$/$d$ quarks, or until we can match to Chiral Perturbation
Theory, $\chi$PT (the problem there being that the radius of 
convergence of $\chi$PT is not known). Also to save CPU time, the fermion
determinant in the action (representing $n_f$ quark flavours)
is often disgarded - the `quenched' approximation.
Finally in addition to all the above problems the matrix element
must be renormalised, in order to be able to compare with the
phenomenological $\overline{MS}$ results.

To attempt to address some of these issues we have generated data sets
(Bakeyev \textit{et al} 2004)
\begin{enumerate}
   \item $O(a)$-improved Wilson fermions (`clover fermions') 
         in the quenched approximation at three
         couplings $\beta \equiv 6/g^2 = 6.0$, $6.2$ and $6.4$
         (G\"ockeler \textit{et al} 2004) corresponding to lattice spacings
         $a^{-1} \sim 2.12$, $2.91$ and $3.85\,\mbox{GeV}$.
         (This checks lattice discretisation errors, which should be
         $O(a^2)$.) The pseudoscalar mass, $m_{ps}$,
         lies between $580\,\mbox{MeV}$ and $1200\,\mbox{MeV}$.
         \label{clover_quenched}
   \item Unquenched clover fermions at $m_{ps}$ down to
         $\sim 560\,\mbox{MeV}$ in order to see if there are any
         discernable quenching effects. Various couplings are used,
         $\beta = 5.20$, $5.25$, $5.29$ and $5.40$ with lattice
         spacings ranging from $a^{-1} \sim 1.61 \,\mbox{GeV}$
         to $2.4\, \mbox{GeV}$.
         \label{clover_unquenched}
   \item Wilson fermions at one fixed lattice spacing,
         $a^{-1} \sim 2.12\,\mbox{GeV}$ in the quenched approximation
         at pseudoscalar masses, $m_{ps}$,  down to $\sim 310\,\mbox{MeV}$,
         ($m_{ps}/m_V \sim 0.4$) in order to try to match
         to chiral perturbation theory. This lattice fermion
         formulation has discretisation errors of $O(a)$.
         \label{wilson_quenched}
   \item Overlap fermions, in the quenched
         approximation at one lattice spacing $a^{-1} \sim 2.09\,\mbox{GeV}$
         down to $m_{ps}$ of about $440\,\mbox{MeV}$.
         These have a chiral symmetry even with finite lattice spacing
         and hence have better chiral properties than either Wilson
         or clover fermions (and also have discretisation errors
         of $O(a^2)$).
\end{enumerate}
Note that the physical pion mass is about $m_\pi \sim 140\,\mbox{MeV}$
and we use the force scale
$r_0 = 0.5\,\mbox{fm} \equiv (394.6\,\mbox{MeV})^{-1}$
to set the scale. These results cover various patches of $(m_{ps}, a, n_f)$
space. This is however not completely satisfactory.
Overlap fermions, although the best formulation of
lattice fermions known, are very expensive in CPU time,
and are only just beginning to be investigated,
eg (Galletly \textit{et al} 2003, Ohta \textit{et al} 2004),
to which we refer the reader to for more details.

The results obtained from eq.~(\ref{Rratio}) are, of course, bare
results and must be renormalised. We shall not discuss this further
here, just noting that many one-loop perturbative results are known;
but are generally not very satisfactory as lattice perturbative
series do not appear to converge very fast. (The convergence
can be helped using tadpole-improvement.) A preferred non-perturbative
method is also available, (Martinelli \textit{et al} 1995) and the
results presented here will have the $Z$s determined by this method.

All our results are for hadrons containing light (ie $u/d$) quarks.
Reaching this limit is extremely costly in CPU time (and except for
the overlap formulation, other problems connected with the
non-chiral nature of the fermion formulation may arise).
Much work has been done recently on chiral perturbation theory
and it would be highly desirable to be in a region
where these results can be matched to lattice results and then
the limit $m_{ps} \to m_\pi$ can be taken.
Although one should take the continuum and chiral limits
separately (and preferably in that order)
we shall try here a variant procedure of using a simultaneous `plane'
fit containing both limits. This is because at present the unquenched,
data set 2 is less complete than the quenched data set 1
and this procedure at least allows for a direct comparison of results.
(For set \ref{clover_quenched} these different fit procedures lead
to similar results.)
Practically we might thus expect that for a quantity $Q$ of interest
\begin{equation}
   Q = F_\chi^Q(r_0m_{ps}) + d_s^Q(a/r_0)^s \,.
\label{Q_fit_general}
\end{equation}
$F_\chi^Q(r_0m_{ps})$ describes the (chiral) physics and the
discretisation errors are $O(a^s)$ where $s=1$ for Wilson fermions
and $s=2$ for clover fermions.
Naively one might expect a Taylor series expansion for $F^Q_\chi$
to be sufficient, ie
\begin{equation}
   F^Q_\chi(x) = Q(0) + c^Qx^2 + \ldots \,,
\label{Q_fit_linear}
\end{equation}
where $x = r_0m_{ps}$. Over the last few years expressions for
$F^Q_\chi$ have been found
\begin{equation}
   F^Q_\chi(x) =  Q(0)\left( 1 -
                  c_{\chi}^Q x^2\ln (x/r_0\Lambda_\chi)^2 \right) + \ldots \,,
\label{Q_fit_chiral}
\end{equation}
showing the existence of a chiral logarithm
$\sim m_q \ln m_q$ (including the quenched case).
For $v_{n;NS}$, $a_{n;NS}$, $t_{n;NS}$ the constant
$c_{\chi}^Q$ is known (and positive), see eg (Chen \textit{et al} 1997).
One expects most effect of the chiral logarithm for $t_{1;NS}$
and least for $a_{0;NS}$. The chiral scale, $\Lambda_\chi$,
is usually taken to be $\sim 1\,\mbox{GeV}$.
The range of validity of the expansion, eq.~(\ref{Q_fit_chiral})
is not known; one might expect that for $m_{ps} > \Lambda_\chi$,
pion loops are suppressed, leading to a smooth variation in $m_q$
ie constituent quarks, while for $m_{ps} < \Lambda_\chi$
non-linear behaviour would be seen. Thus building in some of the
constituent or heavy quark mass expectations, an equation of the form
\begin{equation}
   F^Q_\chi(x) =  Q(0)\left( 1 -
                  c_{\chi}^Q x^2\ln { x^2 \over 
                               (x^2 + (r_0 \Lambda_\chi)^2) }
                               \right) + c^Q x^2  \,,
\label{Q_fit_chiral_practical}
\end{equation}
has been proposed (Detmold \textit{et al} 2001).

Present (numerically) investigated matrix elements include
$v_2 \equiv \langle x \rangle$ (also part of the momentum sum rule:
$\sum_q \langle x \rangle^{(q)} + \langle x \rangle^{(g)} = 1$),
$v_3 \equiv \langle x^2 \rangle$, $v_4 \equiv \langle x^3 \rangle$,
$a_0 = 2\Delta q$ (also occurring in neutron decay, as well as the
Bjorken sum rule, as $\Delta u - \Delta d = g_A$ and connected with
quark spin), $a_1 = 2\Delta q^{(2)}$, $a_2 = 2\Delta q^{(3)}$,
$t_0 = 2\delta q$, $t_1 = 2\delta q^{(2)}$ and $d_2$.
We shall only discuss $v_n$, $n = 1,2,3$,
$a_0/2$, $t_0/2$ and $d_2$ here.


\section{Results of Continuum/Chiral Extrapolations for some twist two
         operators}

\begin{figure}[htb]
   \hspace*{0.50in}
   \epsfxsize=10.00cm 
      \epsfbox{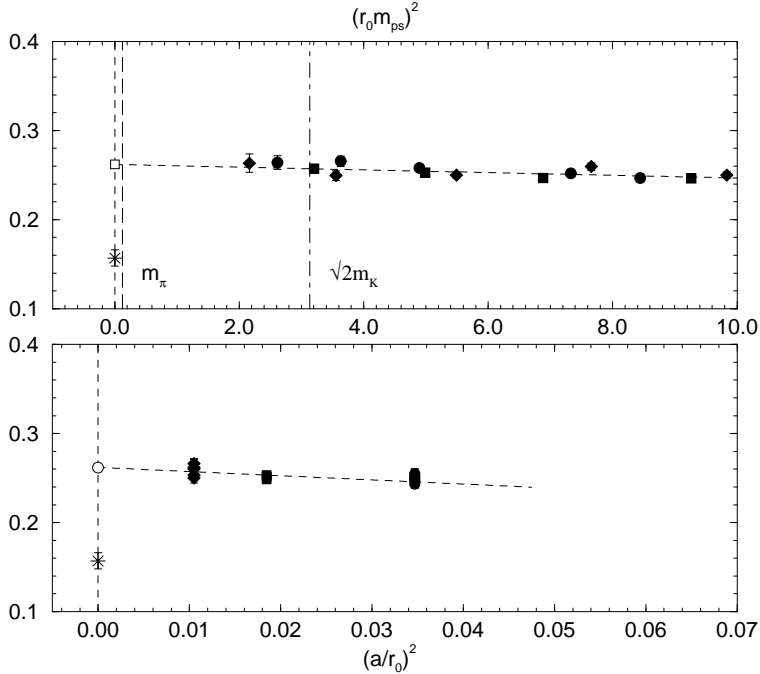}
   \caption{$v_{2;NS}^{\msbar}(2\,\mbox{GeV})$ versus $(r_0m_{ps})^2$
            (upper plot) and versus $(a/r_0)^2$ (lower plot)
            using data set \ref{clover_quenched}.
            Filled circles, squares and diamonds represent the three
            lattice spacings corresponding to $\beta = 6.0$, $6.2$, $6.4$.
            The chiral limit $(r_0m_{ps})^2 = 0$
            is shown as a short-dashed line, while the physical pion
            mass is denoted by the long-dashed line. Also shown as
            a dot-dashed line is the mass of a hypothetical
            $\overline{s}s$ meson calculated as $\sim \sqrt{2}m_K$.
            The MRST phenomenological value is denoted by a star.}
   \label{fig_x1b_1u-1d.p0_040821_1716_standrews04}
\end{figure}

We now show some results. We start by considering
$v_n^{\msbar}(2\,\mbox{GeV})$ for $n = 2$, $3$, $4$.
In Fig.~\ref{fig_x1b_1u-1d.p0_040821_1716_standrews04}
we show $v_{2;NS}^{\msbar}$ from data set 1, together with a fit using
eqs.~(\ref{Q_fit_general}) and (\ref{Q_fit_linear}). We see that 
$O(a^2)$ discretisation errors are small and seem to be relatively
benign. By this we mean that the only limiting factor with the
extrapolation is the amount of data available. We shall (thus)
in future assume that this limit is not a problem.
This does not seem to be the case with the chiral extrapolation
where the data seems to strongly favour a {\it linear} extrapolation
rather than the $\chi$PT result in eq.~(\ref{Q_fit_chiral}).
The value found in the chiral limit is about $50\%$ larger than the
MRST phenomenological value, (Martin \textit{et al} 2002).
(Note however that there are exciting hints that overlap fermions
may be closer to the phenomenological value, (Galletly \textit{et al} 2003,
G\"urtler \textit{et al} 2004).)

\begin{figure}[htb]
   \hspace*{0.50in}
   \epsfxsize=10.00cm 
      \epsfbox{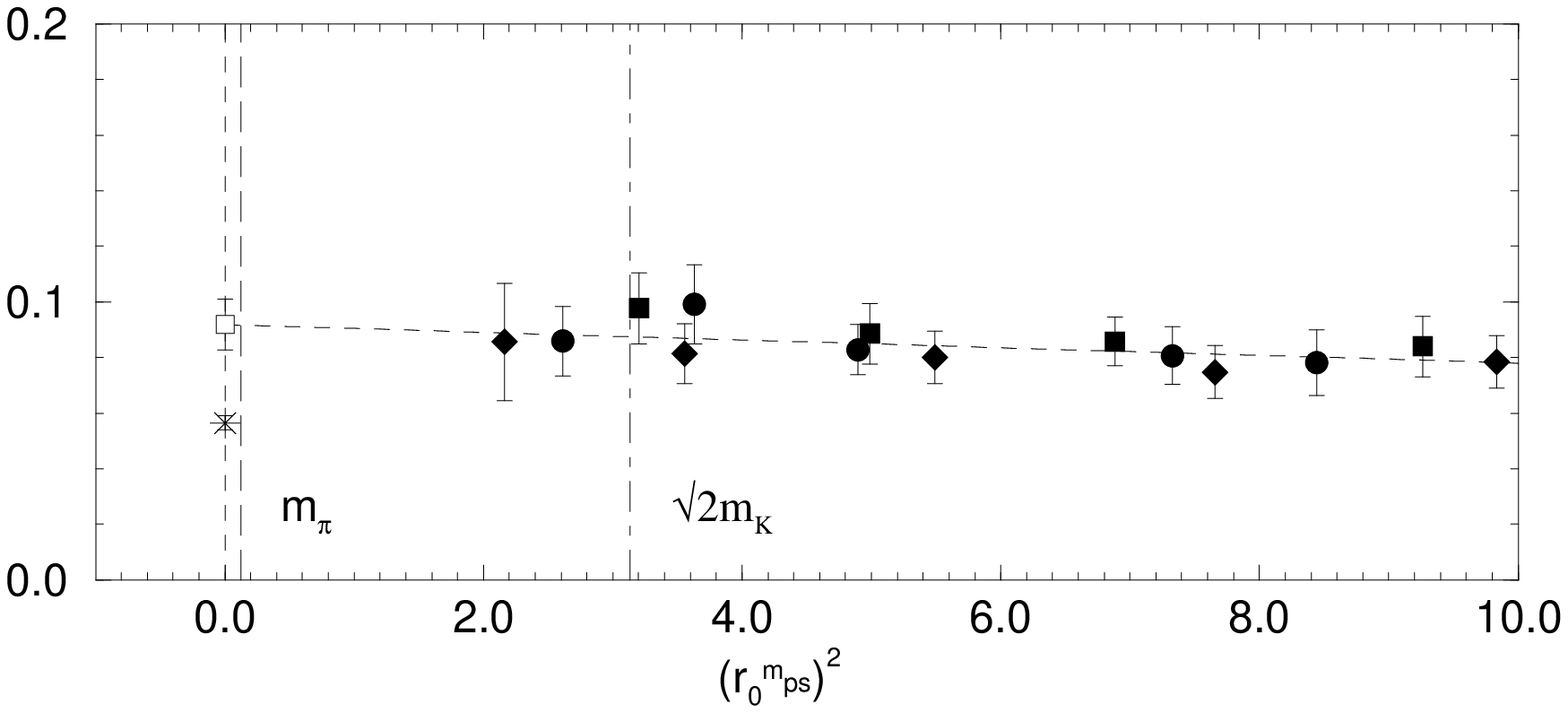}
   \caption{$v_{3;NS}^{\msbar}(2\,\mbox{GeV})$ versus $(r_0m_{ps})^2$.
            Same notation as in 
            Fig.~\ref{fig_x1b_1u-1d.p0_040821_1716_standrews04}.}
   \label{fig_x2b_1u-1d.pm1_040824_1726_standrews04}
\end{figure}

\begin{figure}[htb]
   \hspace*{0.50in}
   \epsfxsize=10.00cm 
      \epsfbox{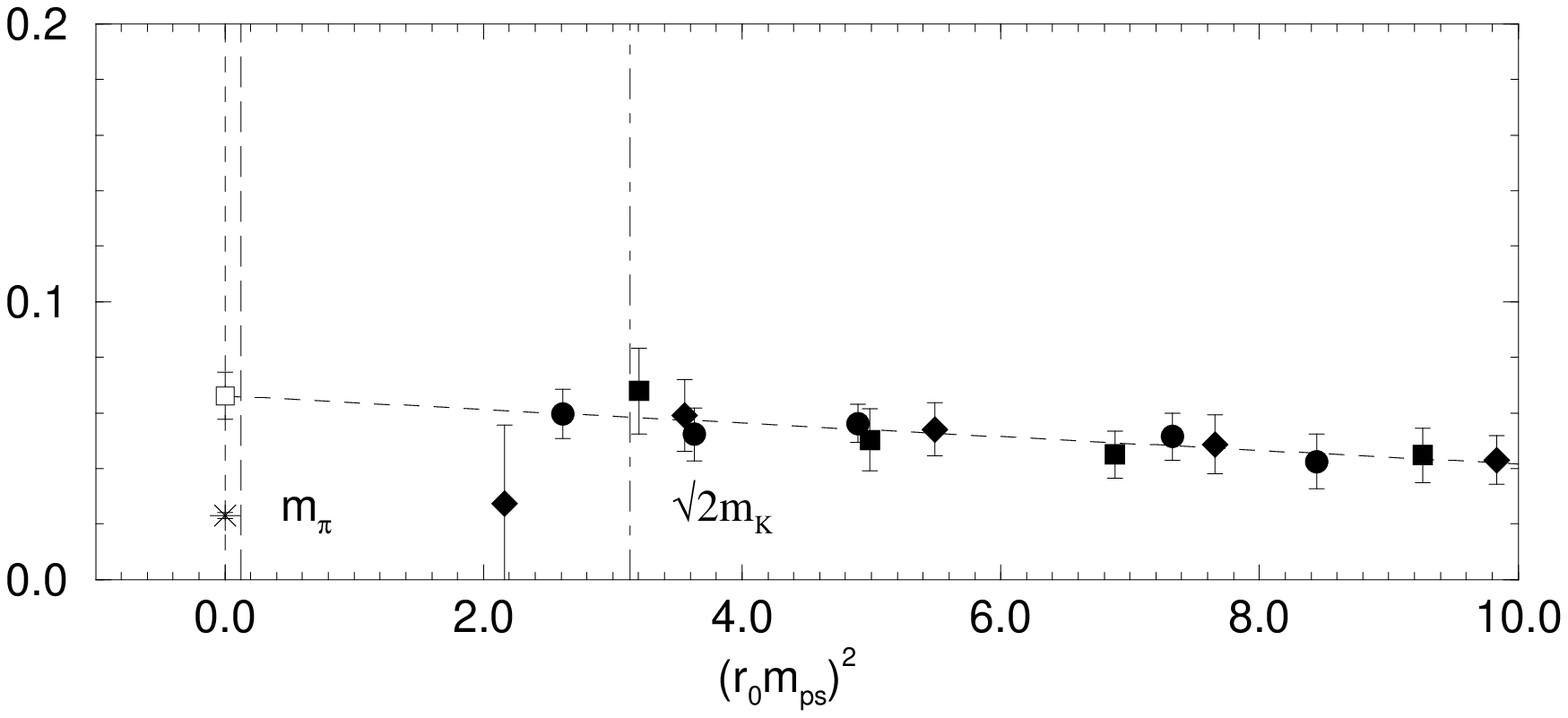}
   \caption{$v_{4;NS}^{\msbar}(2\,\mbox{GeV})$ versus $(r_0m_{ps})^2$.
            Same notation as in 
            Fig.~\ref{fig_x1b_1u-1d.p0_040821_1716_standrews04}.}
   \label{fig_x3b_1u-1d.pm1_040824_1729_standrews04}
\end{figure}

\begin{figure}[htb]
   \hspace*{0.50in}
   \epsfxsize=10.00cm 
      \epsfbox{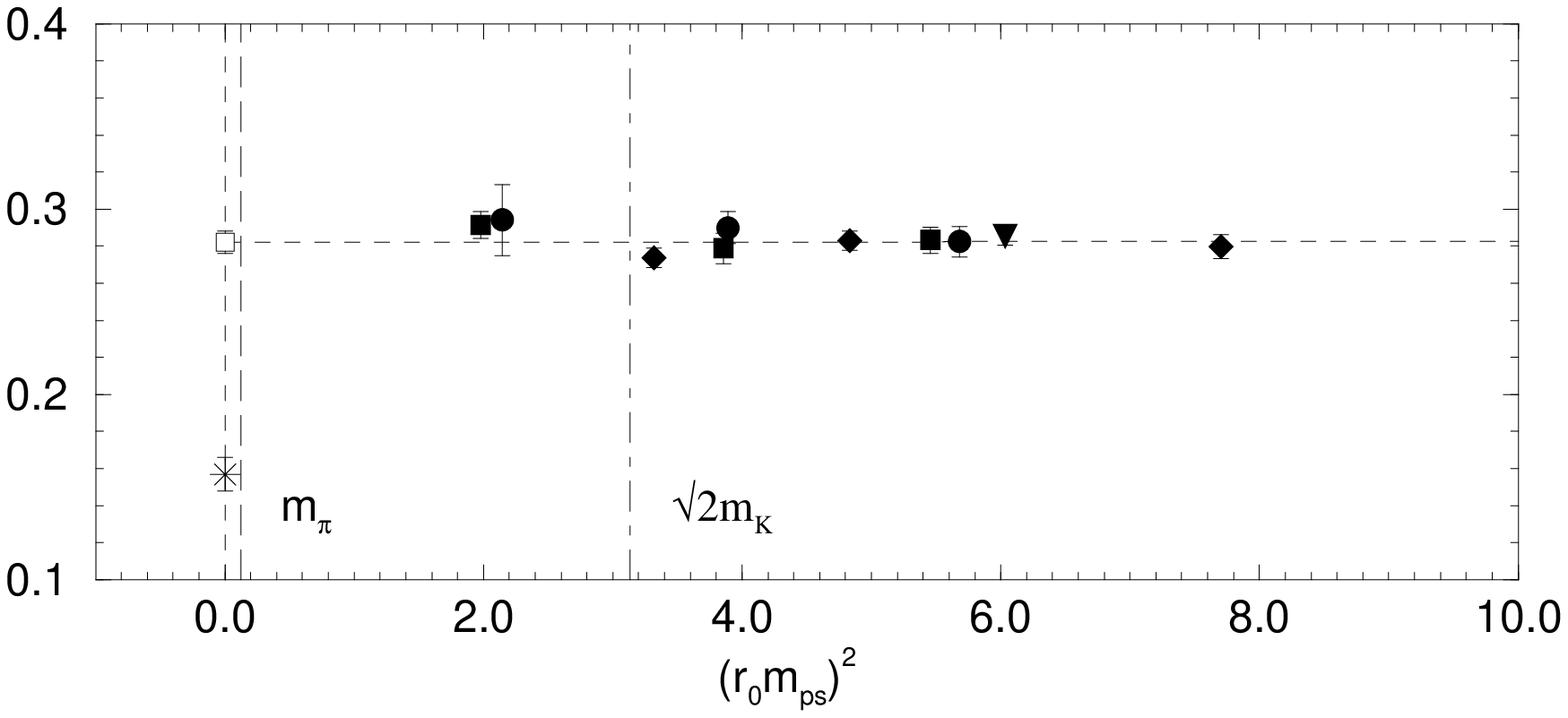}
   \caption{$v_{2;NS}^{\msbar}(2\,\mbox{GeV})$ versus $(r_0m_{ps})^2$
            for unquenched fermions using data set 2.
            $\beta = 5.20$ results are (filled) circles; $5.25$ squares;
            $5.29$, diamonds; $5.40$ down triangle.
            Otherwise the same notation as in 
            Fig.~\ref{fig_x1b_1u-1d.p0_040821_1716_standrews04}.}
   \label{fig_x1b_1u-1d.p0_041124_1113_r0mps2_standrews04}
\end{figure}

The same situation persists for the higher moments $v_3^{\msbar}$
and $v_4^{\msbar}$.
In Figs.~\ref{fig_x2b_1u-1d.pm1_040824_1726_standrews04} and
\ref{fig_x3b_1u-1d.pm1_040824_1729_standrews04} we show
these moments and compare the results with the MRS phenomenological
values. In all cases we find that the moments
are too large in comparison with phenomenological result.
It is not clear why this is so, again the quarks seem to be acting more like
constituent quarks rather than current quarks. Possible causes are
quenching and/or a chiral extrapolation from too heavy a quark mass. 
We first consider possible quenching effects.
In Fig.~\ref{fig_x1b_1u-1d.p0_041124_1113_r0mps2_standrews04}
we consider $v_{2;NS}^{\msbar}$ again, but this time for
$n_f=2$ flavours using data set 2. No real difference is seen
in comparison to the quenched case. Indeed for other matrix elements
considered a similar situation prevails.

\begin{figure}[htb]
   \hspace*{0.50in}
   \epsfxsize=10.00cm 
      \epsfbox{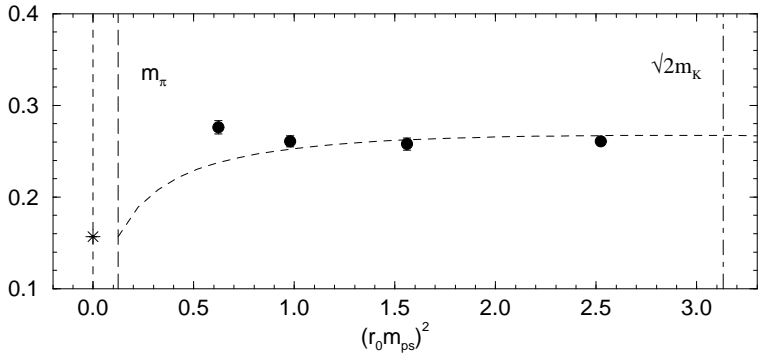}
   \caption{$v_{2;NS}^{\msbar}(2\,\mbox{GeV})$ versus $(r_0m_{ps})^2$
            for Wilson fermions from data set 3. The `fit' uses
            eq.~(\ref{Q_fit_chiral_practical}). Otherwise
            the same notation as in 
            Fig.~\ref{fig_x1b_1u-1d.p0_040821_1716_standrews04}.}
   \label{fig_x1b_1u-1d.p0_040823_1625_standrews04_pap}
\end{figure}

To try to examine the situation at smaller quark mass, we now
turn to the data set 3. Most of the above
results have a quark mass at the strange quark mass (or heavier).
In this data set we have generated quenched Wilson data at one
lattice spacing, at light pion masses down to $310\,\mbox{MeV}$.
In Fig.~\ref{fig_x1b_1u-1d.p0_040823_1625_standrews04_pap}
we show $v_{2;NS}^{\msbar}(2\,\mbox{GeV})$. In comparison with the
previous pictures note that the $x$-scale only runs
to $(r_0m_{ps})^2 \sim 3.0$. Again (except possibly for
the lightest pion mass) the data seems rather linear (and constant).
Also shown is a forced fit from eq.~(\ref{Q_fit_chiral_practical}),
leaving $\Lambda_\chi$ and $c_{\chi}^{v_2}$ free but constrained to go
through the MRST phenomenological value at $m_{ps}=m_\pi$.
Ignoring the lightest quark mass point,
this is just possible; however it is very unnatural giving, for example,
$\Lambda_\chi \sim 500 \,\mbox{GeV}$ which is a very low value.

\begin{figure}[htb]
   \hspace*{0.50in}
   \epsfxsize=10.00cm
      \epsfbox{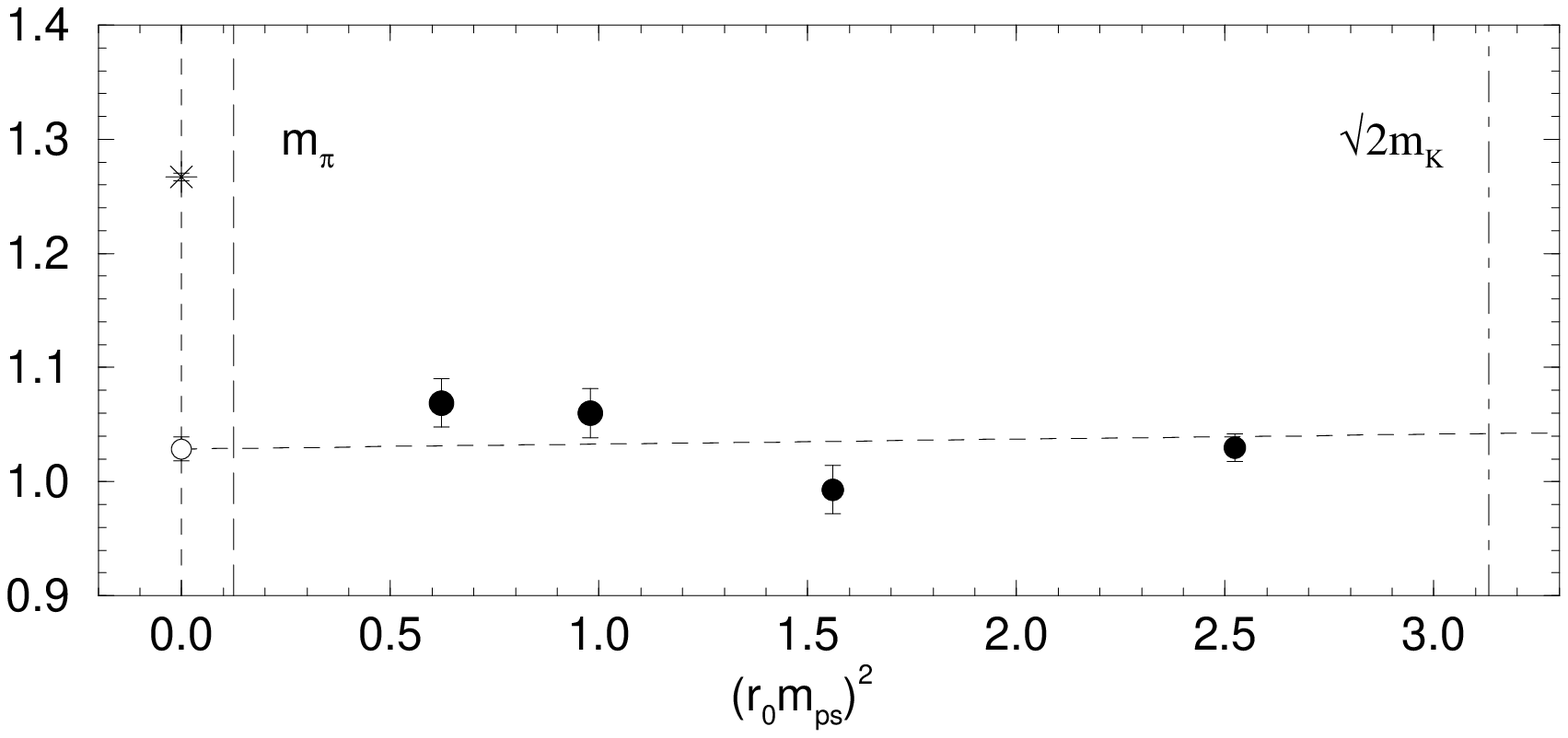}
   \caption{$a_{0;NS}/2 = g_A$ versus $(r_0m_{ps})^2$
            for Wilson fermions (data set 3), with a
            linear fit. Otherwise the same notation as in 
            Fig.~\ref{fig_x1b_1u-1d.p0_040821_1716_standrews04}.}
   \label{fig_axial_2_1u-1d.p0_040819_1150_wil_standrews04}
\end{figure}

\begin{figure}[htb]
   \hspace*{0.50in}
   \epsfxsize=10.00cm
      \epsfbox{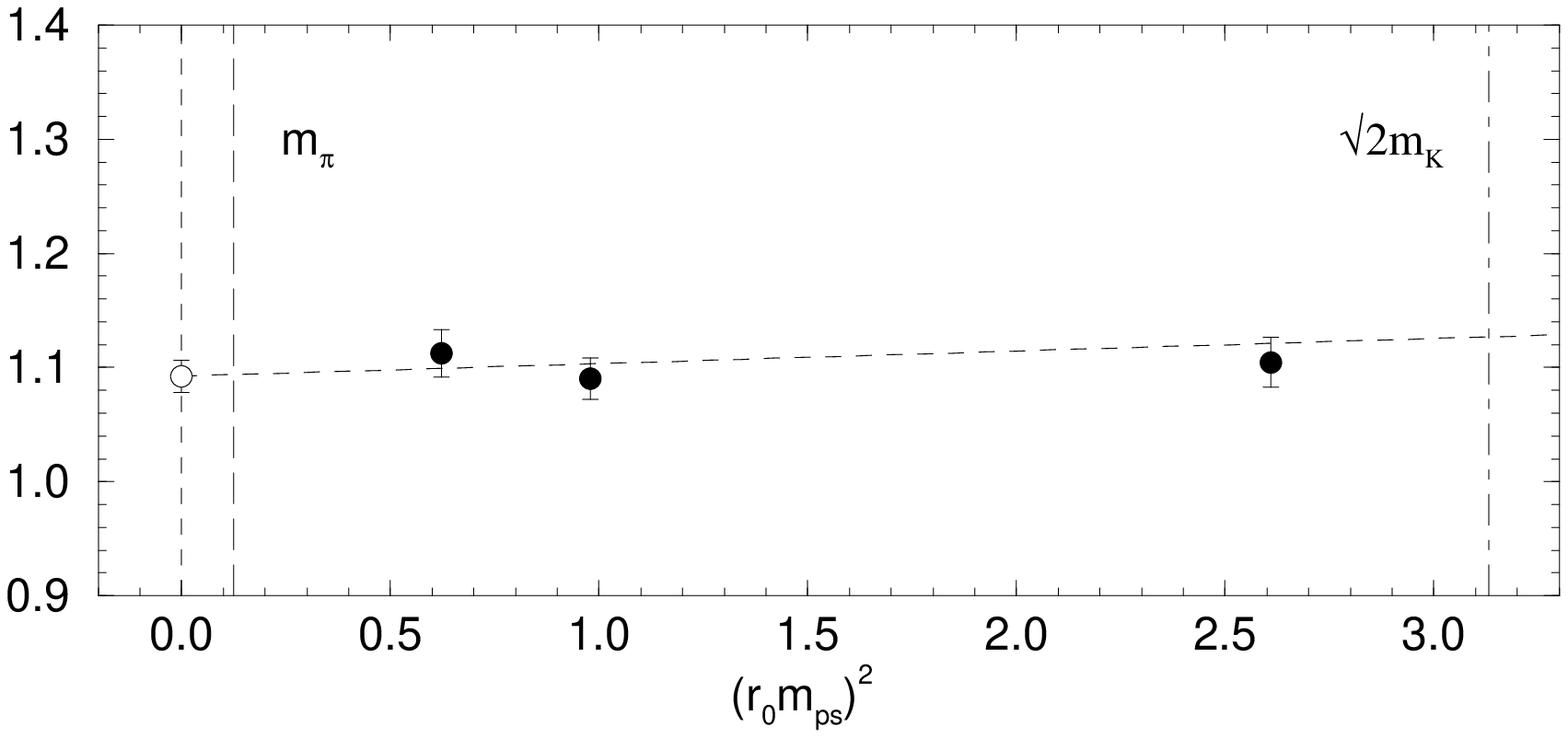}
   \caption{$t_{0;NS}/2$ versus $(r_0m_{ps})^2$
            using data set 3. The same notation as in 
            Fig.~\ref{fig_axial_2_1u-1d.p0_040819_1150_wil_standrews04}.}
   \label{fig_h1b_1u-1d.p0_040820_1806_wil_standrews04}
\end{figure}

A similar situation holds for the axial $a_{0;NS}/2 = g_A$
and tensor charge $t_{0;NS}/2$. In
Figs.~\ref{fig_axial_2_1u-1d.p0_040819_1150_wil_standrews04} and
\ref{fig_h1b_1u-1d.p0_040820_1806_wil_standrews04}.
Again the results in both cases seem very linear. Indeed from
eq.~(\ref{Q_fit_chiral_practical}) due to the negative sign, we must have 
the lattice data decreasing to the phenomenological value.
This is certainly not the case here (although experimentally
$t_{0;NS}$ is not known, we expect a similar situation as for $g_A$;
indeed in the non-relativisitic limit $t_{0;NS}/2 \to g_A$).
Later work including the $\Delta$ as well as the $N$
in chiral perturbation theory (Detmold \textit{et al} 2002,
Hemmert \textit{et al} 2003) reduce the $c_{\chi}^{g_A}$ coefficient,
but this still is a problem. We show only linear fits, giving for $g_A$
a value somewhat lower than the experimental one.
(There are two possible caveats: for clover fermions the continuum
extrapolation may have significant $O(a^2)$ effects, in distinction
to $v_{n;NS}$ and also there may or may not be larger finite volume effects
present both in the data and theoretically, see for example the discussion
in (Cohen 2001).) But at present the same general picture emerges
as for the unpolarised moments.


\section{Some higher twist operator results}


\subsection{Twist three}


The prime example is given by $d_2$, which can 
be determined from $g_2$, the first moment of which is a linear
combination of $a_2$ and $d_2$. The operators for the $a_n$ moments
have twist two, but $d_n$ corresponds to twist three and is thus
of particular interest.
A `straightforward' lattice computation, (G\"ockeler \textit{et al} 1996),
gave rather large values for $d_2^{p}$ (where
$d_2^{(p)} = Q^{(u)2}d_2^{(u)} + Q^{(d)2}d_2^{(d)}$).
A recent experiment, (E155 Collaboration 1999), however indicated
that this term was very small, which would mean that $g_2$ is
almost completely determined by $g_1$ (the Wandzura--Wilczek relation).
This problem was traced in (G\"ockeler \textit{et al} 2000a),
to a mixing of the original operator with a lower-dimensional operator.
This additional operator mixes $\propto 1/a$ and so its
renormalisation constant must be determined non-perturbatively.
In (G\"ockeler \textit{et al} 2000a) this procedure was attempted,
and led to results qualitatively consistent with the experimental values.
Note that this is only a problem when using Wilson or clover like
fermions, as we would expect the additional operator to appear like
$\sim m_q\overline{q} \sigma \Dd{} q$ and hence vanish in the chiral limit.
Thus there should be no mixing if one uses overlap fermions.


\subsection{Twist four}


Potential higher twist effects are present in the moment of a structure
function, see eq.~(\ref{f2sf}).
These $O(1/Q^2)$ terms are composed of dimension 6, four quark matrix
elements. A general problem is the non-perturbative
mixing of these operators with the previous dimension 4 operators.
At present results are restricted to finding combinations
of these higher twist operators which do not mix from flavour symmetry.
For the nucleon the $SU_F(3)$ flavour symmetry group must be considered,
ie taking mass degenerate $u$, $d$ and $s$ quarks,
(G\"ockeler \textit{et al} 2002b) giving
\begin{equation}
   \int_0^1 dx F_2(x,Q^2)|^{27,I=1}_{Nachtmann} =
        -0.0005(5) { m_p^2 \alpha_s(Q^2) \over Q^2 } + O(\alpha_s^2) \,,
\end{equation}
for quenched Wilson fermions (ie part of data set 3).
To access this moment experimentally needs very exotic
combinations of moments from the measurement of the
$p$, $n$, $\Lambda$, $\Sigma$ and $\Xi$ baryons and is not possible.
Nevertheless this term is very small in comparison with the leading
twist result, and might hint that higher twist contributions are small.


\section{Miscellaneous Pion and Lambda results}


Moments for the pion and rho structure functions were 
computed in (Best \textit{et al} 1997), for unimproved Wilson fermions.
Using the Schr\"odinger Functional method, $v_2$ was recently calculated
for the pion, (Guagnelli \textit{et al} 2004)
for both unimproved and $O(a)$-improved fermions.
A higher twist matrix element for the pion has also been computed
for quenched Wilson fermions (ie using part of data set 3)
\begin{equation}
   \int_0^1 dx F_2(x,Q^2)|^{I=2}_{Nachtmann} =
        1.67(64) { f_\pi^2 \alpha_s(Q^2) \over Q^2 } + O(\alpha_s^2) \,,
\end{equation}
where the $SU_F(2)$ flavour symmetry group gives the combination
$F_2^{I=2}=F_2^{\pi^+} + F_2^{\pi^-} -2 F_2^{\pi^0}$. This is again
a rather small number, so although a rather exotic combination of
pion matrix elements it might indicate that higher twist terms are small.

Finally there have been results for moments of $\Lambda$
structure functions, (G\"ockeler \textit{et al} 2002b)
again using Wilson fermions (ie part of data set 3).
These are potentially useful results as one can compare with nucleon
spin structure and check violation of $SU_F(3)$ symmetry. First indications
are that there is no evidence of flavour symmetry breaking
in the matrix elements ie that $\Lambda$ and $p$ are related by
an $SU(3)_F$ flavour transformation.


\section{Conclusions}

Clearly the computation of many matrix elements giving low moments of
structure functions is possible. We would like to emphasise that
a successful computation is a fundamental test of QCD -- this
is not a model computation. There are however many problems to overcome:
finite volume effects, renormalisation and mixing, continuum and
chiral extrapolations and unquenching.
At present although overall impressions are encouraging,
still it is difficult to re-produce experimental/phenomenological 
results of (relatively) simple matrix elements, eg $v_2$.
But progress is being made by the various groups working in the field.
For example in comparison to our previous results,
(G\"ockeler \textit{et al} 1996) there are now non-perturbative $Z$s
and considerations of both chiral and continuum extrapolations
and some unquenched results are now available.
While the continuum extrapolation seems to be `just' a matter of
more data at smaller lattice spacing, the chiral extrapolation
does seem to present a problem, with no sign of any chiral logarithms
being seen as predicted by $\chi$PT. Clearly everything depends on the
data and the quest for better results should continue.
To leave the region where constituent quark masses give a reasonable
description of the data (ie linearity) unfortunately requires
pion masses rather close to the physical pion mass.
In this region fermions with better chiral properties will probably
be needed, such as overlap, which in turn will need much faster machines.


\section*{Acknowledgements}

I wish to thank my co-workers in the QCDSF and UKQCD Collaborations:
A. Ali Khan, T. Bakeyev, D. Galletly,
M. G\"ockeler, M. G\"urtler, P. H\"agler,
T.~R. Hemmert, A.~C. Irving,
B. Jo\'o, A.~D. Kennedy, B. Pendleton,
H. Perlt, D. Pleiter, P.~E.~L. Rakow,
A. Sch\"afer, G. Schierholz, A. Schiller, W. Schroers,
T. Streuer, H. St\"uben, V. Weinberg and J.~M. Zanotti
for a pleasant and profitable collaboration.

The numerical calculations have been performed on the Hitachi SR8000 at
LRZ (Munich), on the Cray T3E at EPCC (Edinburgh)
(Allton \textit{et al} 2002)
on the Cray T3E at NIC (J\"ulich) and ZIB (Berlin),
as well as on the APE1000 and Quadrics (QH2b) at DESY (Zeuthen).
We thank all institutions. This work has been supported in part by
the EU Integrated Infrastructure Initiative Hadron Physics,
contract number RII3-CT-2004-506078
and by the DFG (Forschergruppe Gitter-Hadronen-Ph\"anomenologie).


\section*{References}

\frenchspacing

\begin{small}

\reference{Allton C~R \textit{et al}, 2002,
   \textit{Phys Rev} \vol D65 054502, hep-lat/0107021.}

\reference{Bakeyev T \textit{et al}, 2004, 
   \textit{Nucl Phys Proc Suppl} \vol 128 82, hep-lat/0311017.}

\reference{Best C \textit{et al}, 1997,
   \textit{Phys Rev} \vol D56 2743, hep-lat/970314.}

\reference{Chen J-W \textit{et al}, 1997,
   \textit{Phys Lett} \vol B523 107, hep-ph/0105197.}

\reference{Cohen T~D, 2001,
   \textit{Phys Lett} \vol B529 50, hep-lat/0112014.}

\reference{Detmold W \textit{et al}, 2001,
   \textit{Phys Rev Lett} \vol 87 172001, hep-lat/0103006.}

\reference{Detmold W \textit{et al}, 2002,
   \textit{Phys Rev} \vol D66 054501, hep-lat/0206001.}

\reference{Dolgov D \textit{et al}, 2002,
   \textit{Phys Rev} \vol D66 034506, hep-lat/0201021.}

\reference{E155 Collaboration, 1999,
   \textit{Phys Lett} \vol B458 529, hep-ex/9901006.}

\reference{Galletly D \textit{et al}, 2003,
   \textit{Nucl Phys Proc Suppl} \vol 129 453, hep-lat/0310028.}

\reference{G\"ockeler M \textit{et al}, 1996,
   \textit{Phys Rev} \vol D53 2317, hep-lat/9508004.}

\reference{G\"ockeler M \textit{et al}, 2000a,
   \textit{Phys Rev} \vol D63 074506, hep-lat/0011091.}

\reference{G\"ockeler M \textit{et al}, 2000b,
   \textit{Nucl Phys} \vol B623 287, hep-lat/0103038.}

\reference{G\"ockeler M \textit{et al}, 2002a,
   \textit{Nucl Phys Proc Suppl} \vol B119 398, hep-lat/0209111.}

\reference{G\"ockeler M \textit{et al}, 2002b,
   \textit{Phys Lett} \vol B545 112, hep-lat/0208017.}

\reference{G\"ockeler M \textit{et al}, 2004,
   hep-ph/0410187.}

\reference{Guagnelli M \textit{et al}, 2004,
   hep-lat/0405027.}

\reference{G\"urtler M \textit{et al}, 2004,
   hep-lat/0409164.}

\reference{Hemmert T~R \textit{et al}, 2003,
   \textit{Phys Rev} \vol D68 075009 hep-lat/0303002.}

\reference{Martin A~D \textit{et al}, 2002,
   \textit{Eur Phys J} \vol C23 73, hep-ph/0110215.}
   
\reference{Martinelli G \textit{et al}, 1995,
   \textit{Nucl Phys} \vol B445 81, hep-lat/9411010.}

\reference{Ohta S \textit{et al}, 2004,
   hep-lat/0411008.}

\end{small}

\nonfrenchspacing


\end{document}